\documentclass[journal,twocolumn]{IEEEtran}

\usepackage{graphicx,times,amsmath,amssymb,mathrsfs,cite,stfloats,booktabs,url,multirow, subfigure}
\usepackage[lined,ruled,commentsnumbered]{algorithm2e}

\makeatletter

\newcommand{\Rmnum}[1]{\expandafter\@slowromancap\romannumeral #1@}
\makeatother

\begin{document}
\title{User Identity Protection in EEG-based Brain-Computer Interfaces}

\author{Lubin~Meng, Xue~Jiang, Jian~Huang,~\IEEEmembership{Senior~Member,~IEEE}, Wei~Li, Hanbin~Luo, Dongrui~Wu
\thanks{L.~Meng, X.~Jiang, J.~Huang, W.~Li and D.~Wu are with the Key Laboratory of Image Processing and Intelligent Control, School of Artificial Intelligence and Automation, Huazhong University of Science and Technology, Wuhan 430074 China. They are also with the Belt and Road Joint Laboratory on Measurement and Control Technology, Huazhong University of Science and Technology, Wuhan 430074, China}
\thanks{H.~Luo is with the School of Civil and Hydraulic Engineering, Huazhong University of Science and Technology, Wuhan, China.}

\thanks{Corresponding author: D.~Wu (e-mail: drwu09@gmail.com)}}

\maketitle

\begin{abstract}
A brain-computer interface (BCI) establishes a direct communication pathway between the brain and an external device. Electroencephalogram (EEG) is the most popular input signal in BCIs, due to its convenience and low cost. Most research on EEG-based BCIs focuses on the accurate decoding of EEG signals; however, EEG signals also contain rich private information, e.g., user identity, emotion, and so on, which should be protected. This paper first exposes a serious privacy problem in  EEG-based BCIs, i.e., the user identity in EEG data can be easily learned so that different sessions of EEG data from the same user can be associated together to more reliably mine private information. To address this issue, we further propose two approaches to convert the original EEG data into identity-unlearnable EEG data, i.e., removing the user identity information while maintaining the good performance on the primary BCI task. Experiments on seven EEG datasets from five different BCI paradigms showed that on average the generated identity-unlearnable EEG data can reduce the user identification accuracy from 70.01\% to at most 21.36\%, greatly facilitating user privacy protection in EEG-based BCIs.
\end{abstract}


\IEEEpeerreviewmaketitle

\section{Introduction}

A brain-computer interface (BCI)\cite{Wolpaw2002} establishes a direct communication pathway between the brain and an external device (computer, wheelchair, robot, etc). It has been used in neurological rehabilitation\cite{Daly2008}, active tactile exploration\cite{Doherty2011}, robotic device control\cite{Hochberg2012,Edelman2019}, awareness evaluation\cite{Li2016a}, speech synthesis\cite{Anumanchipalli2019}, cortical activity to text translation\cite{Makin2020}, and so on. Electroencephalogram (EEG), which measures the brain signals from the scalp, is the most popular input of BCIs, due to its convenience and low cost.

Machine learning has been extensively used to recognize complex EEG patterns in different BCI paradigms, such as motor imagery (MI)\cite{drwuMITLBCI2022}, event-related potentials (ERP)\cite{Hueebner2018}, emotion recognition\cite{Shanechi2019}, etc. Usually, adequate EEG data are needed to train an accurate machine learning model. However, EEG data contain not only task-specific information but also rich private information\cite{drwuTCSS2023}. For example, Martinovic \emph{et al.}\cite{Martinovic2012} showed that various private information (credit cards, PIN numbers, known people, residential addresses) could be inferred from EEG signals. Icena \textit{et al.}\cite{Ienca2018} pointed out that direct-to-consumer neuro-technologies, including neuro-monitoring headsets and neuro-modulation tools, may all have ethical concerns including privacy. Choi \emph{et al.}\cite{Choi2018} demonstrated that resting-state EEG signals could be used to infer the user identity at 88.4\% accuracy. Landau \emph{et al.}\cite{Landau2020} also showed that meaningful personality traits and cognitive abilities could be predicted by analyzing resting-state EEG recordings.

To accommodate increasing privacy concerns, multiple laws worldwide, e.g., the General Data Protection Regulation of the European Union and the Personal Information Protection Law of China, have been established to enforce strict user privacy protection. As a result, in addition to simple anonymization and data sanitization, various more sophisticated privacy-protection approaches have been proposed for EEG-based BCIs, which can be grouped into the following two categories\cite{drwuTCSS2023}:
\begin{enumerate}
\item Cryptography, which typically includes homomorphic encryption, secure multi-party computation, and secure processors. For example, Agarwal \emph{et al.}\cite{drwuPrivacy2019} proposed secure multiparty computation based cryptographic protocols for privacy protection in EEG-based driver drowsiness estimation.

\item Privacy-preserving machine learning, which performs machine learning without seeing the raw EEG data or model parameters. Typical approaches include federated learning and source-free transfer learning. For example, Gu \emph{et al.}\cite{Gu2022} proposed frame-level teacher-student learning for EEG-based emotion recognition, where the user data privacy is protected by passing only the model weights instead of the EEG data to the student network. Xia \emph{et al.}\cite{Xia2022} proposed augmentation-based source-free adaptation, which performs privacy-preserving transfer learning in MI classification without accessing the source EEG data, or even the source model parameters. Zhang and Wu\cite{drwuLSFT2022} solved the same problem using lightweight source-free transfer. Zhang \emph{et al.}\cite{drwuMSDT2022} further proposed unsupervised privacy-preserving multi-source decentralized transfer and demonstrated its effectiveness in MI classification and EEG-based emotion classification. Li \emph{et al.}\cite{drwuMeta2022} proposed multi-domain model-agnostic meta-learning for privacy-preserving cross-subject and few-shot MI/ERP classification.
\end{enumerate}

\begin{figure*}[htpb]\centering
\includegraphics[width=0.9\linewidth,clip]{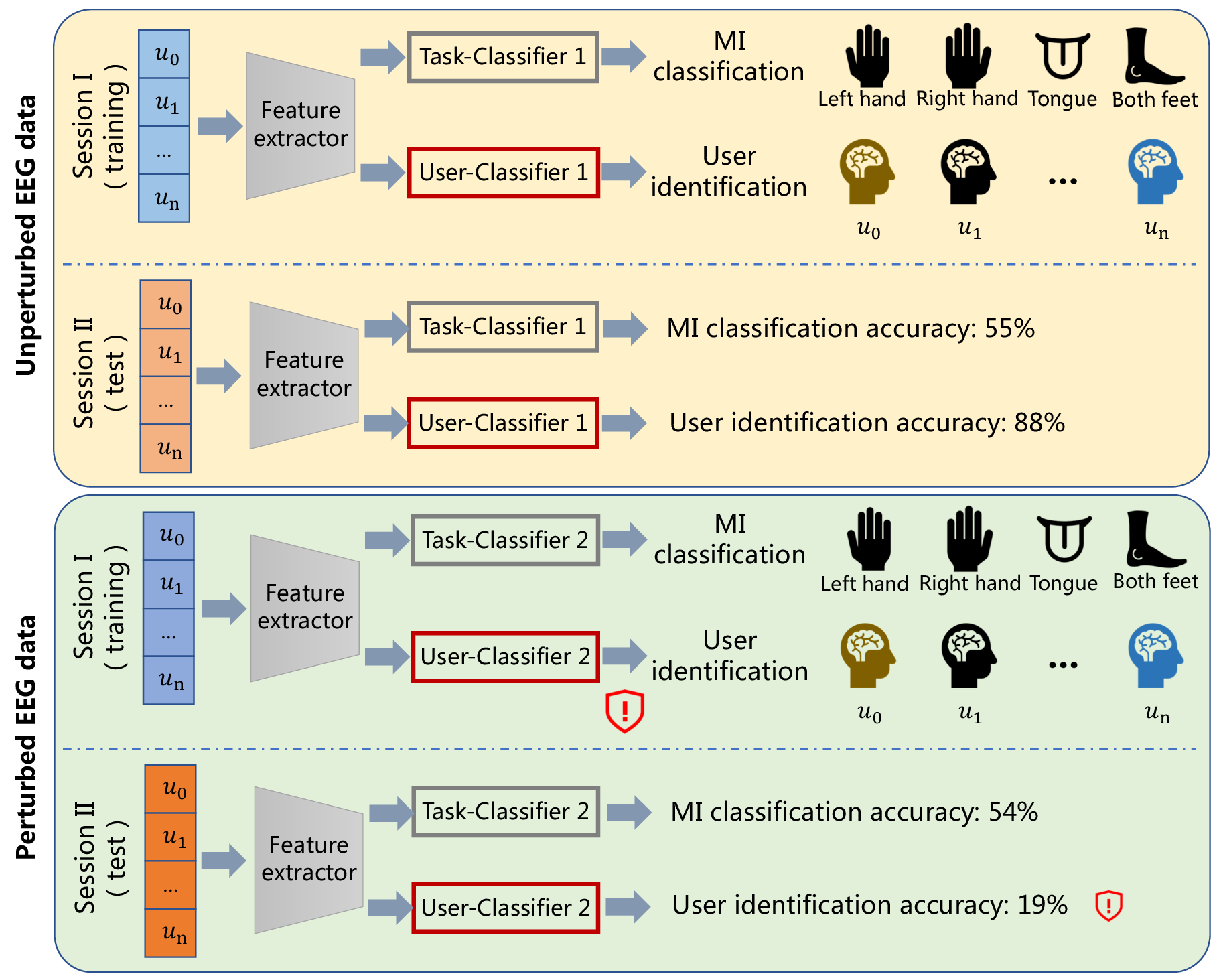}
\caption{Illustration of user identity protection in EEG-based MI classification. The EEG data of Session II can be accurately associated with the unperturbed EEG data of Session I from the same user while performing the primary MI classification task. For the perturbed EEG data, the high MI classification accuracy is maintained, whereas the test session cannot be associated with the training session from the same user.} \label{fig:fig1}
\end{figure*}

Another popular privacy-protection strategy is applying perturbation, which adds perturbation to or transforms the original data to hide private information while maintaining their utility for downstream tasks. Typical approaches include differential privacy and data reconstruction. However, to our knowledge, this strategy has not been investigated in EEG-based BCIs. This paper fills this gap by proposing a machine learning based perturbation approach to hide user identities in EEG-based BCIs. We first demonstrate that EEG data for different BCI tasks can be used to determine the user's identity, and then propose a framework to convert the original EEG data into identity-unlearnable ones while preserving the task-specific information for the primary BCI task.

Hiding user identities is very important in EEG-based BCIs. Consider the following scenario: a company collects EEG data from multiple volunteers, each with multiple sessions, to train a reliable MI classifier for its products. When a volunteer is performing the MI task, private health information like depression or seizure may be detected from the EEG signals, and the detection accuracy increases with the amount of EEG data. This privacy concern may hinder the recruitment of volunteers. However, if the user identity information can be reliably removed, i.e., multiple sessions from the same user are perturbed so that they cannot be associated with the same individual, then the risk of privacy leakage is significantly reduced, and volunteers may be more willing to contribute their data.

Another related example is EEG-based seizure classification. Consider the scenario that a company is collecting seizure EEG data from multiple hospitals to train an accurate and robust classifier. Each patient may have multiple sessions of EEG recordings during the diagnostics and treatment process in a hospital. If an algorithm can find out which sessions belong to the same patient, then it is possible to further infer how effective the treatment from that hospital is. This is sensitive information that a hospital may not want to disclose, as it affects its competitiveness and reputation. If we can remove the patient identities so that it is impossible to know which sessions are from the same patient, then more hospitals may be willing to participate.

Simple anonymization cannot remove the user identity completely. Fig.~\ref{fig:fig1} uses MI-based BCIs (to distinguish the imagined movement of the left hand, right hand, both feet, and tongue, from EEGs) as an example. Task-Classifier 1 and User-Classifier 1 are trained on the unperturbed EEG data for MI classification and user classification, respectively. Task-Classifier 2 and User-Classifier 2, which have the same structures as Task-Classifier 1 and User-Classifier 1, respectively, are trained on the perturbed identity-unlearnable EEG data. In the test phase, both Task-Classifiers achieve similar MI classification accuracies (55\% versus 54\%); however, User-Classifier 1 can also associate the test data with the unperturbed training data from the same user accurately (88\%), whereas User-Classifier 2 has poor performance (19\%). Thus, the perturbed identity-unlearnable training EEG data protect the user identity privacy while maintaining a similar performance in the primary MI classification task.

Our main contributions are:
\begin{enumerate}
	\item We demonstrate that various BCI paradigms can leak user identity information, hence the necessity of user identity protection.
	\item We propose two approaches to generate identity-unlearnable EEG data, which can be used to protect the user identity privacy while maintaining the primary task classification accuracy in five different BCI paradigms.
\end{enumerate}

\section{Method}

This section introduces the details to the generation of identity-unlearnable EEG data.

\subsection{Problem statement}

Given an EEG dataset $\mathcal{D}=\{(\mathbf{x}_i, y_i, u_i)\}_{i=1}^N$, where $\mathbf{x}_i \in \mathcal{X} \subset \mathbb{R}^{\mathrm{c} \times \mathrm{t}}$ is the $i$-th EEG trial with $\mathrm{c}$ channels and $\mathrm{t}$ time domain samples, $y_i \in \mathcal{Y}=\{1,\dots,K\}$ the task label (e.g., left fist and right fist in MI1) and $u_i \in \mathcal{U}=\{1,\dots,U\}$ the user identity label of the $i$-th EEG trial, $N$ the number of EEG trails.

A feature extractor $F$ and a Task-Classifier $C$ can be trained on $\mathcal{D}$ to learn task-related patterns. For brevity, we denote the model as $H_C=C \circ F$, which maps the input space into a label space, i.e., $H_C: \mathcal{X} \rightarrow \mathcal{Y}$.

However, in addition to the task-related information, EEG data may also contain rich private information, such as the user identity. Although $F$ is primarily used to extract task-related features, the extracted features also contain rich user identity information. So, a User-Classifier $D$ can be constructed to learn the user identity information from the features extracted from $F$, i.e., $H_D=D \circ F: \mathcal{X} \rightarrow \mathcal{U}$ maps the input space into a user-identity space.

This paper aims to make the identity information in EEG data unlearnable, while preserving the task-related information. We generate an identity-unlearnable EEG dataset $\mathcal{D}'=\{(\mathbf{x}'_i, y_i, u_i)\}_{i=1}^N$, where $\mathbf{x}'=\mathbf{x}+\boldsymbol{\delta}$ and $\boldsymbol{\delta}$ is a specifically designed perturbation. This dataset can be used to train $H_C$ normally as the original unperturbed dataset $\mathcal{D}$, but it is difficult to train $H_D$ to learn the user identity information.

\subsection{Sample-wise perturbation generation}

Inspired by Reference\cite{Huang2021}, we add a crafted perturbation to \emph{each EEG trial} that may have a strong correlation with the user identity, misleading the machine learning model to learn the perturbation pattern rather than the true user identity pattern. Simultaneously, we minimize the impact of the perturbation on the task-related information, ensuring its normal use in the primary task.

Specifically, we first train a feature extractor $F'$, a Task-Classifier $C'$, and a User-Classifier $D'$, to evaluate the impact of the perturbation on the task-related information and the identity-related information, which may be different from $F$, $C$ and $D$. Denote the models as $H'_C=C' \circ F'$ and $H'_D=D' \circ F'$. The loss function is
\begin{align}
	\min_{\boldsymbol{\theta}_{H'_C}, \boldsymbol{\theta}_{H'_D}} \mathbb{E}_{(\mathbf{x},y,u) \sim \mathcal{D}}\left[\ell_{\mathrm{CE}}\left(H'_C(\mathbf{x}), y\right) + \alpha \ell_{\mathrm{CE}}\left(H'_D(\mathrm{x}), u\right)\right], \label{eq: objective_model}
\end{align}
where $\ell_{\mathrm{CE}}$ is the cross-entropy loss, and $\alpha$ a trade-off parameter.

Then, given an EEG trial $\mathbf{x}$, we generate the corresponding perturbation $\boldsymbol{\delta}$ by
\begin{align}
	\min_{\boldsymbol{\delta}} \ & \ell_\mathrm{MSE}\left(H'_C(\mathbf{x}+\boldsymbol{\delta}), H'_C(\mathbf{x})\right) \nonumber \\ & + \beta \ell_\mathrm{CE}\left(H'_D(\mathbf{x}+\boldsymbol{\delta}), u\right) \quad \mathrm{s.t.} \quad \|\boldsymbol{\delta}\|_{\infty} \leq \epsilon , \label{eq: objective}
\end{align}
where $\ell_\mathrm{MSE}$ is the mean squared error, $\beta$ a trade-off parameter, and $\epsilon$ the maximum perturbation amplitude. The first term constrains the impact of $\boldsymbol{\delta}$ on the task-related information, and the second term enhances the correlation between the perturbation $\boldsymbol{\delta}$ and the user identity $u$. A first-order optimization method, project gradient descent\cite{Madry2018}, can be used to solve this minimization problem iteratively:
\begin{align}
	\boldsymbol{\delta}_0 &= \boldsymbol{\xi}, \label{eq: initial} \\
	\boldsymbol{\delta}_i &=\mathrm{Proj}_{0, \epsilon}\left(\boldsymbol{\delta}_{i-1} - \eta \mathrm{sign}\left( \bigtriangledown_{\mathbf{x}+\boldsymbol{\delta}_{i-1}} \mathcal{L}(\mathbf{x},\boldsymbol{\delta}_{i-1},u)\right) \right), \label{eq: iteration}
\end{align}
where $i=1,\cdots,n_{\mathrm{iter}}$, and $n_{\mathrm{iter}}$ is the number of iterations. $\boldsymbol{\xi}\in \mathrm{Uniform}(-\epsilon, \epsilon)$ is a randomly initialized perturbation, $\eta$ the gradient step size, and $\mathcal{L}$ the loss function in~(\ref{eq: objective}). $\mathrm{Proj}_{0, \epsilon}(\cdot)$ projects its input onto the $\ell_{\infty}$ ball of radius $\epsilon$ centered at 0.

To improve the effectiveness of the perturbation $\boldsymbol{\delta}$, we update $\boldsymbol{\delta}$ after multiple (instead of only one) epochs of model training. Specifically, we first train the models by~(\ref{eq: objective_model}) for $L$ epochs, then update $\boldsymbol{\delta}$ with them by~(\ref{eq: iteration}). After generating the perturbations for all EEG trials on dataset $\mathcal{D}$, we train the models on dataset $\mathcal{D}'=\{(\mathbf{x}_i+\boldsymbol{\delta}_i, y_i, u_i)\}_{i=1}^N$ again. The perturbation updates repeat $M$ times, and the final dataset $\mathcal{D}'$ is the identity-unlearnable EEG dataset.

The pseudo-code of sample-wise perturbation generation is given in Algorithm~\ref{alg:sample-wise}.

\begin{algorithm}[!t] 
\KwIn{$\mathcal{D}=\{(\mathbf{x}_i, y_i, u_i)\}_{i=1}^N$, the original EEG dataset\;
\hspace*{11mm}$H'_C$, the task classifier\;
\hspace*{11mm}$H'_D$, the user classifier\;
\hspace*{11mm}$L$, the number of model training epochs before perturbation update\;
\hspace*{11mm}$M$, the maximum number of perturbation updates\;
\hspace*{11mm}$\boldsymbol{\xi}$, a randomly initialized perturbation\;}
\KwOut{An identity-unlearnable EEG dataset $\mathcal{D}'=\{(\mathbf{x}'_i,y_i,u_i)\}_{i=1}^N$.}

Initialize $\boldsymbol{\delta} \leftarrow \boldsymbol{\xi}$, $\mathcal{D}' \leftarrow \mathcal{D}$\;
\For{$m=1,\cdots,M$}{
	\For{$l=1,\cdots,L$}{
		Update $H'_C$ and $H'_D$ by~(\ref{eq: objective_model}) on $\mathcal{D}'$\;
	}
	Update $\boldsymbol{\delta}$ by~(\ref{eq: objective}) on $\mathcal{D}$\;
	Update $\mathcal{D}'=\{(\mathbf{x}_i+\boldsymbol{\delta}_i,y_i,u_i)\}_{i=1}^N$\;
}

\textbf{Return} $\mathcal{D}'$
\caption{Sample-wise perturbation generation.} \label{alg:sample-wise}
\end{algorithm}

\subsection{User-wise perturbation generation}

To speed up perturbation generation, we further propose a user-wise perturbation generation approach to design a universal perturbation template for \emph{each user}. Compared with the sample-wise perturbation, the user-wise perturbation is more convenient to implement in practice. One only needs to generate $U$ perturbations to make the identity information of all EEG trials unlearnable.

To generate the user-wise perturbation, we also first train models $H'_C$ and $H'_D$ by~(\ref{eq: objective_model}) on dataset $\mathcal{D}$ till convergence. Then, we optimize $\boldsymbol{\delta}=[\boldsymbol{\delta}_1,\cdots,\boldsymbol{\delta}_U]$ by gradient descent:
\begin{align}
	\min_{\boldsymbol{\delta}} \mathbb{E}_{(\mathbf{x},y,u)\sim \mathcal{D}} \big[
	     &\ell_\mathrm{MSE}\left(H'_C(\mathbf{x}+\boldsymbol{\delta}_u), H'_C(\mathbf{x})\right)  \nonumber \\
	     & + \beta \ell_\mathrm{CE}\left(H'_D(\mathbf{x}+\boldsymbol{\delta}_u),u \right)  \nonumber \\
         & + \gamma \|\boldsymbol{\delta}_u\|_2 \big], \label{eq: user-wise}
\end{align}
where $\boldsymbol{\delta}_u$ is the perturbation for user $u$, $\beta$ the trade-off parameter, and $\gamma$ the regularization coefficient. The first two terms are similar to those in~(\ref{eq: objective}). The last term constrains the perturbation amplitude.

The pseudo-code of user-wise perturbation generation is described in Algorithm~\ref{alg:user-wise}.

\begin{algorithm}[htpb]
\KwIn{$\mathcal{D}=\{(\mathbf{x}_i, y_i, u_i)\}_{i=1}^N$, the original EEG dataset\;
\hspace*{11mm}$H'_C$, the task classifier\;
\hspace*{11mm}$H'_D$, the user classifier\;
\hspace*{11mm}$M_{\mathrm{model}}$, the maximum number of model training epochs\;
\hspace*{11mm}$M_{\mathrm{perturbation}}$, the maximum number of perturbation optimization epochs\;}
\KwOut{An identity-unlearnable EEG dataset $\mathcal{D}'=\{(\mathbf{x}'_i,y_i,u_i)\}_{i=1}^N$.}
Initialize $\boldsymbol{\delta} \leftarrow \mathcal{N}(0,0.001)$\;
\For{$m=1,\cdots,M_{\mathrm{model}}$}{
    Update $H'_C$ and $H'_D$ by~(\ref{eq: objective_model}) on $\mathcal{D}$\;
}
\For{$m=1,\cdots,M_{\mathrm{perturbation}}$}{
    Update $\boldsymbol{\delta}$ by~(\ref{eq: user-wise}) on $\mathcal{D}$\;
}

\textbf{Return} $\mathcal{D}'=\{(\mathbf{x}_i+\boldsymbol{\delta}_{u_i},y_i,u_i)\}_{i=1}^N$.
\caption{User-wise perturbation generation.} \label{alg:user-wise}
\end{algorithm}

\section{Experimental Settings}

This section introduces the experimental settings for validating the effectiveness of the identity-unlearnable EEG data.

\subsection{Datasets}

This study used the following five publicly available EEG datasets:
\begin{enumerate}
\item Motor imagery 1 (MI1)\cite{Schalk2004}: This MI dataset was collected from 109 users. Each user performed four MI tasks, while 64-channel EEG data were recorded. Task 2 (imagine the opening and closing of the left or right fist) was used in this paper, which contains 45 EEG trials from three sessions for each user. For preprocessing, we applied a [4, 40]Hz band-pass filter and down-sampled the EEG data to 128Hz. We then extracted EEG trials in [0.5, 2.5]s after each MI task.
\item Motor imagery 2 (MI2)\cite{Tangermann2012}: This MI dataset is Dataset 2a in BCI Competition IV. It consists of 22-channel EEG data from 9 users who performed four different MI tasks (left hand, right hand, both feet, and tongue), each with 144 trials from two sessions. The same preprocessing pipeline as that on MI1 was used.
\item P300 evoked potentials (P300)\cite{Hoffmann2008}: The P300 dataset was collected from 8 users. Each user had about 3,300 EEG trials with two classes (target and non-target) from four sessions. For preprocessing, we filtered the 32-channel EEG data with a [1, 40]Hz band-pass filter, down-sampled them to 128Hz, and extracted trials in [0, 1]s after each image onset.
\item Feedback error-related negativity (ERN)\cite{ERN}: The ERN dataset was used in a competition at the 2015 IEEE Neural Engineering Conference, hosted on Kaggle. The competition provided a training set from 16 users and a test set from 10 users. The training set was used in this paper. Each user had 340 trials with two classes (good-feedback and bad-feedback) from five sessions. For preprocessing, the 56-channel EEG data were band-pass filtered to [1, 40]Hz, down-sampled to 128Hz, and each trial was extracted in [0, 1.3]s after each stimulus onset.
\item Steady-state visual evoked potential (SSVEP)\cite{Wang2017}: The SSVEP dataset was acquired using a 40-target BCI speller. It consists of 64-channel EEG data from 35 users, each with six blocks. Each block had 40 trials, one for each target. For preprocessing, we applied a [6, 90]Hz band-pass filter and down-sampled the EEG data to 128Hz, and extracted EEG trials in [1, 5]s after each stimulus onset.
\item Neonatal seizures (NS)\cite{Stevenson2019}: The NS dataset contains 79 users, each with about 74-minute 19-channel EEG recordings. The EEG data were labeled by three human experts with one-second resolution. 14 seizure patients were used in this paper, 95\% of whose EEG labels were consistent from the three experts. For preprocessing, we band-pass filtered the EEG data to [0.5, 70]Hz, down-sampled them to 128Hz, and then segmented them into non-overlapping 1-second trials.
\item The Temple University Hospital EEG Seizure Corpus (TUSZ)\cite{Obeid2016}: The TUSZ dataset contains a training set from 579 users, an evaluation set from 43 users, and a development set from 53 users. 292 users in the training set, whose EEG recordings were longer than four seconds, were used in this study. The same preprocessing pipeline as that on NS was used. Additionally, since there are two types of EEGs (average re-referenced and linked ears re-referenced), we converted them to a temporal central parasagittal montage\cite{Lopez2016}.
\end{enumerate}

\subsection{Models}

We used the following three CNN models without the last fully-connected layer as feature extractors:
\begin{enumerate}
	\item EEGNet\cite{Lawhern2018}: EEGNet is a compact CNN architecture specifically designed for EEG-based BCIs. It consists of two convolutional blocks, where depthwise and separable convolutions instead of traditional convolutions are used to reduce the number of model parameters.
	\item DeepCNN\cite{Schirrmeister2017}: DeepCNN contains four convolutional blocks. The first convolutional block is specifically designed for EEG inputs, and the other three are standard convolutional blocks.
	\item ShallowCNN\cite{Schirrmeister2017}: ShallowCNN is a shallow version of DeepCNN, inspired by filter bank common spatial patterns. It has only one convolutional block with a larger kernel and a different pooling approach, compared with DeepCNN.
\end{enumerate}
Details of the architectures of EEGNet, DeepCNN, and ShallowCNN are given in Supplementary Tables 1-3, respectively.

We used one fully-connected layer as the Task-Classifier, and two fully-connected layers as the User-Classifier.

\subsection{Experimental settings}

For each dataset, we performed leave-one-session-out cross-validation (i.e., one session as the training set, and the remaining sessions as the test set). Specifically, assume that the dataset contains $S$ sessions; then, one of the $S$ sessions was used to train a Task-Classifier and a User-Classifier, and the remaining $S-1$ sessions were used to evaluate the classification accuracy. This validation process was repeated $S$ times so that each session became the training set once. The entire cross-validation process was repeated 5 times with different random seeds. The average BCAs and UIAs of the $5S$ runs were reported.

Since the NS dataset has only one session, we divided each user's data into three equal-length sessions. In the TUSZ dataset, since different users had varying numbers of sessions, we used the first session of each user and divided it into two equal-length sub-sessions.

\subsection{Performance measures}

As there is significantly class imbalance in some BCI tasks, balanced classification accuracy (BCA) was used to evaluate the Task-Classifier's performance:
\begin{align} \mathrm{BCA}=\frac{1}{K}\sum_{k=1}^K\frac{1}{N_k}\sum_{i=1}^{N_k}\mathbf{1}(y_{\mathrm{pred},i}=k),
\end{align}
where $K$ is the number of classes, $N_k$ the number of test samples in class $k$, $y_{\mathrm{pred},i}$ the classifier's prediction on the $i$-th test sample, and $\mathbf{1}(\cdot)$ an indicator function.

To measure the difficulty of mining the user identity information from the EEG data, the user identification accuracy (UIA) was used to evaluate the User-Classifier's performance:
\begin{align} \mathrm{UIA}=\frac{1}{N_t}\sum_{i=1}^{N_t}\mathbf{1}(u_{\mathrm{pred},i}=u_i),
\end{align}
where $N_t$ is the number of test samples, $u_{\mathrm{pred},i}$ the classifier's prediction on the $i$-th test sample, and $u_i$ the true user identity.

\subsection{Hyper-parameters}

The hyper-parameters for generating the sample-wise and user-wise perturbations in our experiments are shown in Tables~\ref{tab:sample-wise} and~\ref{tab:user-wise}, respectively.

\begin{table}[htpb]
\centering
\caption{Hyper-parameters for sample-wise perturbation generation.} \renewcommand\arraystretch{1.3}\setlength{\tabcolsep}{5.0mm}
\begin{tabular}{c|c}
\toprule
Parameter & Value \\ \midrule
$\alpha$ & $0.1$ \\
\multirow{2}{*}{$\beta$}  & $0.03/0.01/1.0/0.5/0.05/0.2/0.05$  \\
                          & for MI1/MI2/P300/ERN/SSVEP/NS/TUSZ \\
$\epsilon$ & $0.01$ \\
$n_{\mathrm{iter}}$ & $5$ \\
$\eta$ & $0.002$ \\
$L$    & $5$     \\
$M$    & $30$    \\ \bottomrule
\end{tabular}
\label{tab:sample-wise}
\end{table}

\begin{table}[htpb]
\centering
\caption{Hyper-parameters for user-wise perturbation generation.} \renewcommand\arraystretch{1.3}\setlength{\tabcolsep}{5.0mm}
\begin{tabular}{c|c}
\toprule
Parameter & Value \\ \midrule
$\alpha$ & $0.1$ \\
\multirow{2}{*}{$\beta$}  & $0.03/0.01/1.0/0.5/0.05/0.2/0.05$  \\
                          & for MI1/MI2/P300/ERN/SSVEP/NS/TUSZ \\
$\gamma$ & $\frac{1E-6}{\beta}$ \\
$M_{\mathrm{model}}$    & $150$     \\
$M_{\mathrm{perturbation}}$    & $150$    \\ \bottomrule
\end{tabular}
\label{tab:user-wise}
\end{table}

\section{Results}

This section validates the effectiveness of the identity-unlearnable EEG data.

\subsection{User identity discovery in EEG-based BCIs}

In addition to the primary task, EEG data also contain rich user identity information.

To demonstrate that, we first trained a feature extractor and a Task-Classifier on the unperturbed training EEG data to perform the primary BCI task (e.g., MI classification), then used the learned task-related feature to train a User-Classifier to discriminate the identity. The test results are shown in the `Unperturbed EEG' panel of Table~\ref{tab:main_result}. Balanced classification accuracies (BCAs) of different BCI tasks were reasonable, and user identification accuracies (UIAs) were much higher than random guess (UIAs on MI1 were relatively low because its number of users is much larger), no matter which convolutional neural network (CNN) model (EEGNet, DeepCNN, or ShallowCNN) was used as the feature extractor. These results confirmed our conjecture: rich user identity information can be exploited from EEG data in different BCI paradigms, so that different sessions of EEG data from the same user can be accurately associated together.

\subsection{User identity protection on the perturbed EEG data}

To protect the user identity information in EEG-based BCIs, we perturbed the original EEG data into identity-unlearnable EEG data, from which machine learning models cannot easily identify which sessions are from the same user.

We generated sample-wise perturbations and user-wise perturbations (EEGNet was used as the feature extractor) and added them to the original EEG data, making the user identity information unlearnable. To evaluate its effectiveness, we first trained a feature extractor and a Task-Classifier on the perturbed training EEG data, then used the extracted features to train a User-Classifier. The results are shown in Table~\ref{tab:main_result}. We can observe that:
\begin{enumerate}
\item The BCAs after both sample-wise perturbations and user-wise perturbations were very close to their counterparts on the unperturbed EEG, indicating that the perturbations had almost no negative impact on the primary BCI tasks.

\item The UIAs after both perturbations were significantly lower than their counterparts on the unperturbed EEG, indicating that little user identity information can be learned from the perturbed EEG data.
\end{enumerate}
Although EEGNet was always used as the feature extractor in training, the learned perturbations were still effective when DeepCNN and ShallowCNN were used as the feature extractor in testing, suggesting the generalization ability of sample-wise and user-wise perturbations. This makes our approach very easy to implement in practice: the user can use an arbitrary neural network model to perturb the EEG data for identity protection.

\begin{table*}[!t] \centering \setlength{\tabcolsep}{4.5mm}
\caption{BCAs and UIAs (\%) on the unperturbed EEG data and those with sample-wise or user-wise perturbations. }
\begin{tabular}{c|c|c|cc|cc|cc|cc}
\toprule
\multirow{2}{*}{Dataset} & Number & Feature & \multicolumn{2}{c|}{Unperturbed} & \multicolumn{2}{c|}{Sample-wise} & \multicolumn{2}{c|}{User-wise} & \multicolumn{2}{c}{Average} \\
 	& of Users      & Extractor  &\multicolumn{2}{c|}{EEG} & \multicolumn{2}{c|}{Perturbation} & \multicolumn{2}{c|}{Perturbation} & \multicolumn{2}{c}{Reduction} \\ \cline{4-11}

 &&& BCA   & UIA    & BCA   & UIA    & BCA   & UIA  & BCA   & UIA\\ \midrule
 \multirow{3}{*}{MI1} & \multirow{3}{*}{109} & EEGNet & 68.30  & 51.48  & 66.83  & 2.30  & 67.47  & 5.90  & 1.15  & 47.38 \\
    & & DeepCNN & 63.54  & 23.95  & 63.05  & 5.54  & 62.29  & 23.48  & 0.87  & 9.44 \\
    & & ShallowCNN & 66.18  & 67.02  & 64.92  & 2.50  & 65.79  & 15.60  & 0.83  & 57.97 \\ \midrule
\multirow{3}{*}{MI2} & \multirow{3}{*}{9} & EEGNet & 57.55  & 67.12  & 51.99  & 40.24  & 52.00  & 15.71  & 5.56  & 39.14 \\
    & & DeepCNN & 53.49  & 72.84  & 53.44  & 48.75  & 53.63  & 48.60  & -0.04  & 24.17 \\
    & & ShallowCNN & 54.81  & 87.89  & 52.75  & 32.56  & 53.86  & 18.94  & 1.51  & 62.14\\  \midrule
\multirow{3}{*}{P300} & \multirow{3}{*}{8} & EEGNet & 67.20  & 81.96  & 66.62  & 32.95  & 65.68  & 19.87  & 1.05  & 55.55 \\
    & & DeepCNN & 66.20  & 82.80  & 65.68  & 42.12  & 64.22  & 25.10  & 1.25  & 49.19 \\
    & & ShallowCNN & 64.61  & 87.98  & 64.14  & 42.53  & 62.91  & 28.95  & 1.08  & 52.24  \\\midrule
\multirow{3}{*}{ERN} & \multirow{3}{*}{16} & EEGNet & 66.68  & 56.52  & 65.05  & 12.32  & 63.70  & 14.37  & 2.30  & 43.17 \\
    & & DeepCNN & 65.43  & 65.84  & 64.41  & 18.36  & 63.34  & 31.93  & 1.55  & 40.69 \\
    & & ShallowCNN & 66.96  & 70.52  & 64.82  & 11.96  & 63.66  & 17.17  & 2.72  & 55.96 \\\midrule
\multirow{3}{*}{SSVEP} & \multirow{3}{*}{35} & EEGNet & 54.04   & 86.89 & 51.96 & 13.36 & 53.61 & 12.88 & 1.26 & 73.77 \\
    & & DeepCNN & 86.67 	& 59.04 & 86.39 & 12.86 & 86.78 & 27.86 & 0.09 & 38.69  \\
    & & ShallowCNN & 55.92 & 98.46  & 43.94 & 7.46  & 55.84 & 6.40  & 6.03 & 91.52  \\\midrule
\multirow{3}{*}{NS} & \multirow{3}{*}{14} & EEGNet & 65.36  & 94.40  & 62.17  & 34.37  & 58.94  & 18.03  & 4.81  & 68.20 \\
    & & DeepCNN & 57.89  & 83.07  & 56.81  & 33.91  & 55.04  & 7.87  & 1.96  & 62.18 \\
    & & ShallowCNN & 43.77 & 93.99 & 33.55 & 8.07 & 42.61 & 10.63 & 5.69 & 84.64  \\\midrule
\multirow{3}{*}{TUSZ} & \multirow{3}{*}{292} & EEGNet & 74.94 & 47.76  & 66.42 & 2.39  & 70.55 & 18.76  & 6.46 & 37.18 \\
    & & DeepCNN & 74.85 & 46.94 & 70.66 & 6.11  & 71.76 & 10.43 & 3.65 & 38.67 \\
    & & ShallowCNN & 74.80 & 47.65 & 68.90 & 2.73 & 70.86 & 18.11 & 4.93 & 37.23 \\\midrule
\multicolumn{3}{c|}{Average}  & 64.67 & 70.01 & 62.11 & 21.36 & 62.35 & 19.64 & 2.43 & 49.51  \\   \bottomrule
\end{tabular}
\label{tab:main_result}
\end{table*}

\subsection{Characteristics of the perturbed EEG data}

We further demonstrate that the identity-unlearnable EEG data are very similar to the original unperturbed EEG data from various perspectives.

\begin{figure}[htbp]\centering
\subfigure[]{\label{fig:fig2a}  \includegraphics[width=1.0\linewidth,clip]{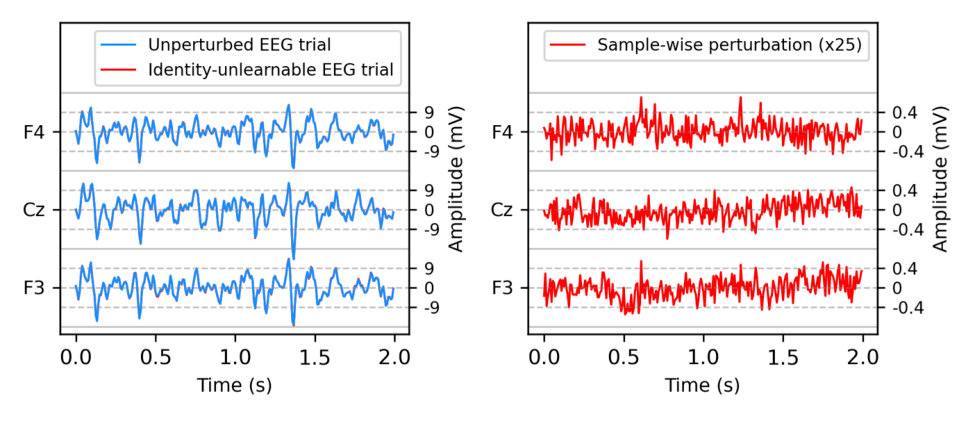}}
\subfigure[]{\label{fig:fig2b}  \includegraphics[width=1.0\linewidth,clip]{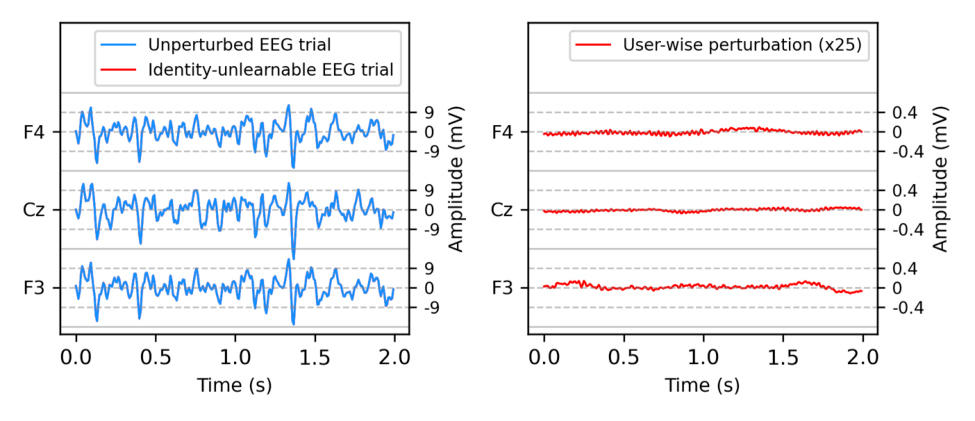}}
\caption{An unperturbed EEG trial and its perturbed counterpart on MI1, which are almost identical. (a) Sample-wise; and, (b) User-wise. The perturbations are magnified $25$ times for better visualization. } \label{fig:2}
\end{figure}

Figs.~\ref{fig:fig2a} and~\ref{fig:fig2b} visualize an original EEG trial from the MI1 dataset, and its two perturbed counterparts. For clarity, only three EEG channels (F4, Cz, and F3) are shown. Similar visualizations on the other six datasets are shown in Supplementary Figs. 1-6. The perturbed EEG trials were always almost identical to the unperturbed ones, regardless of the BCI task. User-wise perturbations on NS in Supplementary Fig. 5 had relatively large amplitudes, which is reasonable, because seizure and normal EEG trials have large magnitude differences, and hence the perturbations should also be larger.

\begin{figure*}[!t]\centering
\subfigure[]{\label{fig:fig3a}  \includegraphics[width=1.0\linewidth,clip]{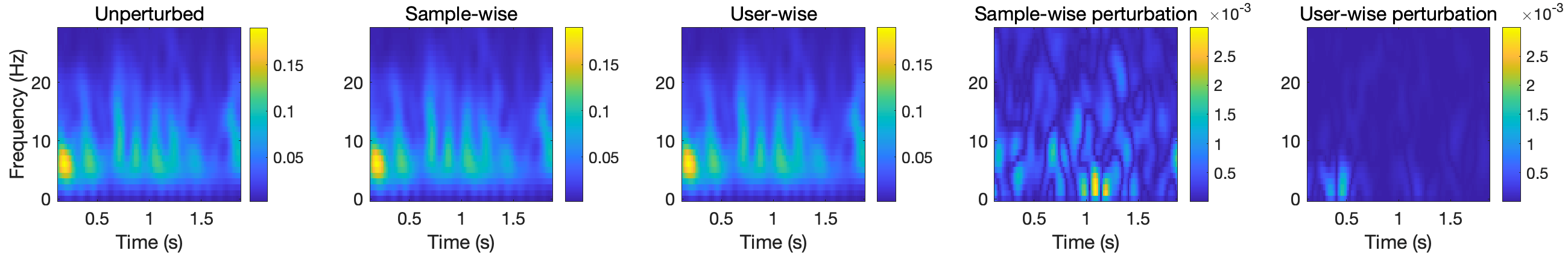}}
\subfigure[]{\label{fig:fig3b}  \includegraphics[width=1.0\linewidth,clip]{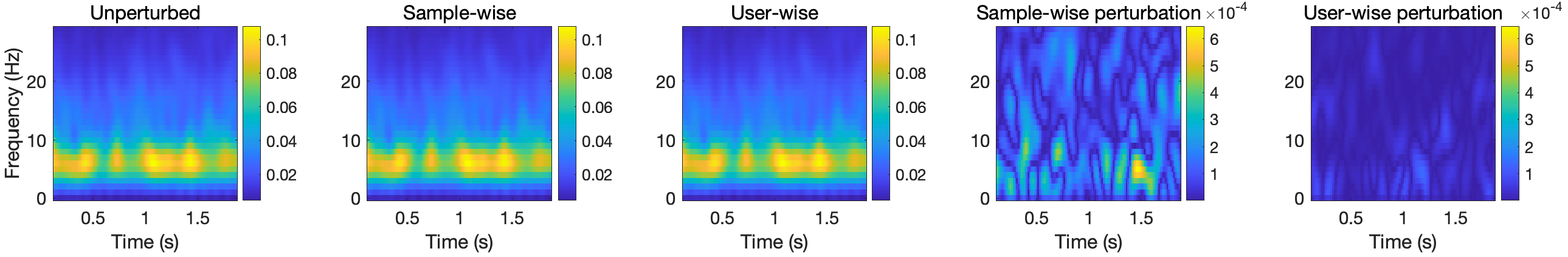}}
\caption{Average Cz channel spectrograms of the unperturbed EEG trials, their perturbed counterparts, and the corresponding perturbations, for the imagination of (a) the left fist on MI1; and, (b) the left hand on MI2.} \label{fig:fig3}
\end{figure*}

Figs.~\ref{fig:fig3a} and~\ref{fig:fig3b} show the average Cz channel spectrograms of the unperturbed EEG trials, their perturbed counterparts, and the corresponding perturbations, when imagining the movement of the left fist on MI1 and the left hand on MI2, respectively. The average spectrograms on the other five datasets are shown in Supplementary Figs.~7-11. One can hardly observe any differences between the unperturbed and perturbed spectrograms.

\begin{figure*}[!t]\centering
\subfigure[]{\label{fig:fig4a}  \includegraphics[width=1.0\linewidth,clip]{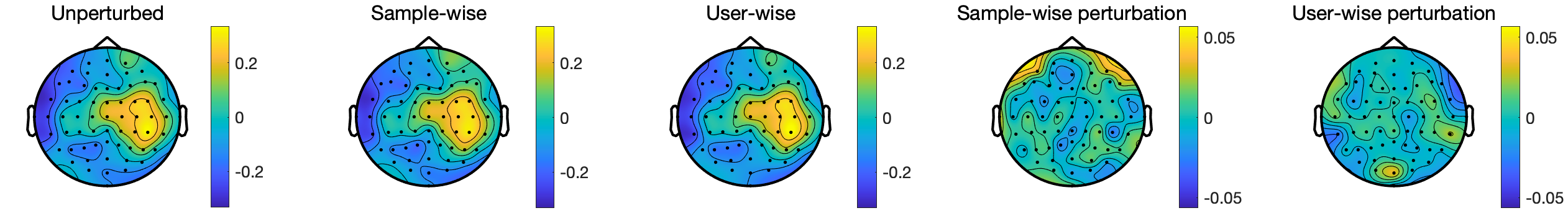}}
\subfigure[]{\label{fig:fig4b}  \includegraphics[width=1.0\linewidth,clip]{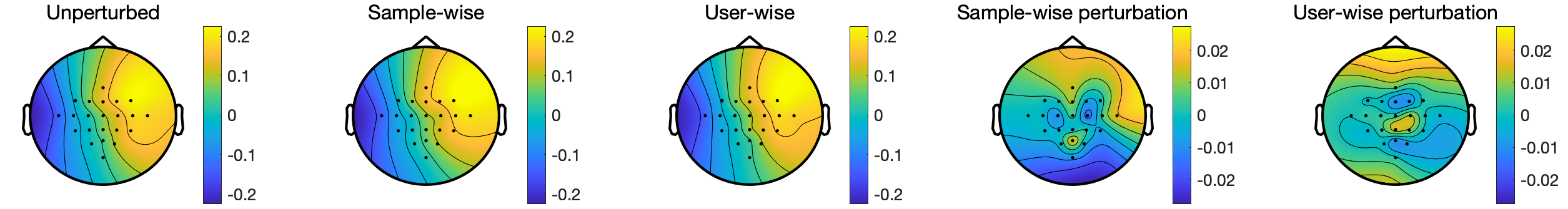}}
\caption{Average topoplots of the unperturbed EEG trials, their perturbed counterparts, and the corresponding perturbations, for the imagination of (a) the left fist on MI1; and, (b) the left hand on MI2.} \label{fig:fig4}
\end{figure*}

Figs.~\ref{fig:fig4a} and~\ref{fig:fig4b} show the average topoplots of the unperturbed EEG trials, their perturbed counterparts, and the corresponding perturbations for the imagination of the left fist on MI1 and the left hand on MI2, respectively. The average topoplots on the other five datasets are shown in Supplementary Figs.~12-16. The topoplots from the unperturbed trials and the perturbed ones are almost identical.

\begin{figure*}[htbp]\centering
\subfigure[]{\label{fig:fig5a}  \includegraphics[width=0.65\linewidth,clip]{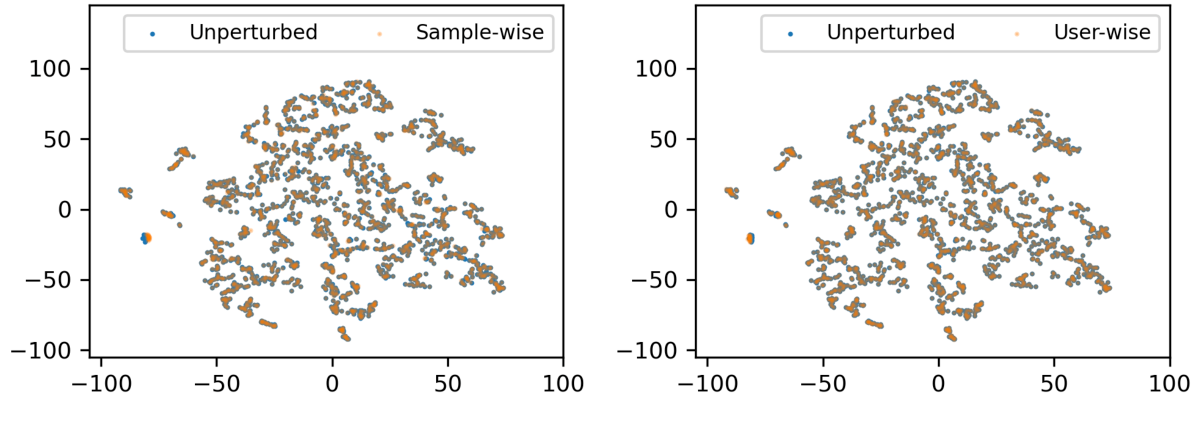}}
\subfigure[]{\label{fig:fig5b}  \includegraphics[width=0.33\linewidth,clip]{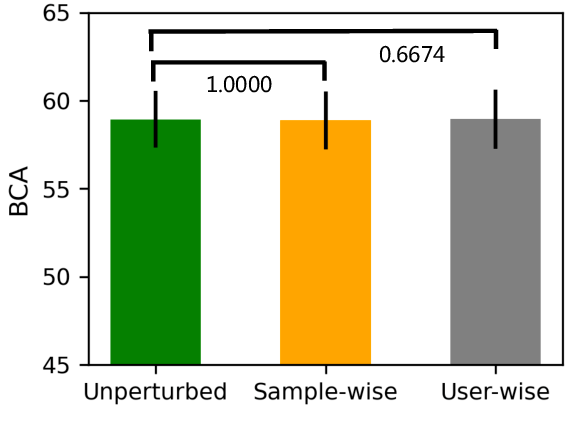}}
\caption{(a) $t$-SNE visualization of common spatial pattern features extracted from the unperturbed EEG trials and their perturbed counterparts on MI1; and, (b) BCAs of logistic regression classifiers trained with common spatial pattern features.} \label{fig:5}
\end{figure*}

\begin{figure*}[htbp]\centering
\subfigure[]{\label{fig:fig6a}  \includegraphics[width=0.66\linewidth,clip]{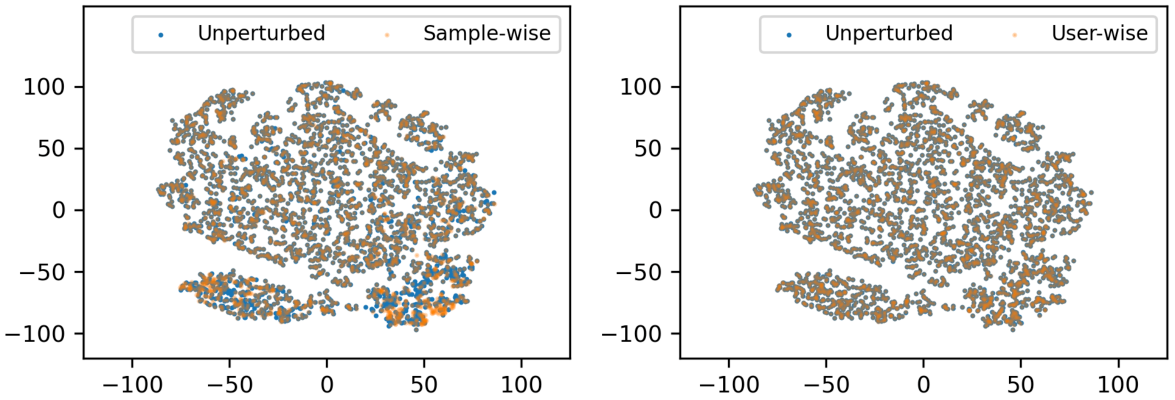}}
\subfigure[]{\label{fig:fig6b}  \includegraphics[width=0.32\linewidth,clip]{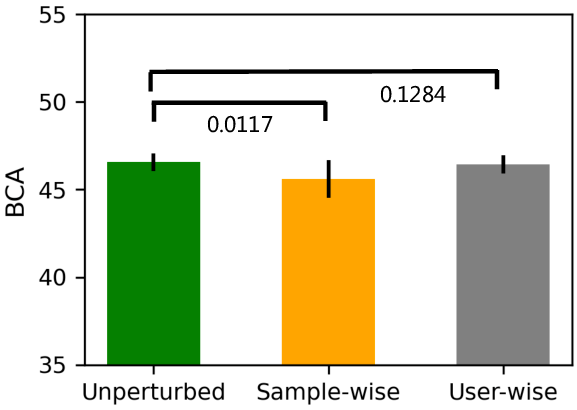}}
\caption{(a) $t$-SNE visualization of common spatial pattern features extracted from the unperturbed EEG trials and their perturbed counterparts on MI2; and, (b) BCAs of logistic regression classifiers trained with common spatial pattern features.} \label{fig:6}
\end{figure*}

In addition to deep learning, manual feature extraction and traditional machine learning models are also frequently used in EEG-based BCIs. Figs.~\ref{fig:fig5a} and~\ref{fig:fig6a} show $t$-SNE\cite{Maaten2008} visualizations of common spatial pattern \cite{drwuMITLBCI2022} features extracted from the unperturbed EEG trials, sample-wise perturbed EEG trials (left), and user-wise perturbed EEG trials (right), on MI1 and MI2, respectively. The overall feature distributions are almost identical, with or without perturbations.

Figs.~\ref{fig:fig5b} and~\ref{fig:fig6b} show BCAs of logistic regression classifiers trained with common spatial pattern features on MI1 and MI2, respectively. To check if the BCA differences between the unperturbed EEG trials and their perturbed counterparts were statistically significant, \emph{p}-values were calculated by the standard \emph{t}-test and adjusted by Benjamini Hochberg False Discovery Rate correction\cite{Benjamini1995}. In most cases, there was no statistically significant difference between the results of the unperturbed EEG trials and the perturbed ones.

All above results demonstrated that the perturbed EEG trials have almost identical characteristics to their unperturbed counterparts, in the primary BCI tasks.

\subsection{Imperceptibility during training}

The high similarity between the identity-unlearnable EEG data and the original unperturbed EEG data can prevent privacy attackers from discovering that the data have been perturbed. We further compared the training processes of Task-Classifier and User-Classifier on the original unperturbed data and the identity-unlearnable data, to validate if the perturbation can be discovered from the training process.

\begin{figure*}[!t]\centering
\subfigure[]{\label{fig:fig7a}  \includegraphics[width=0.8\linewidth,clip]{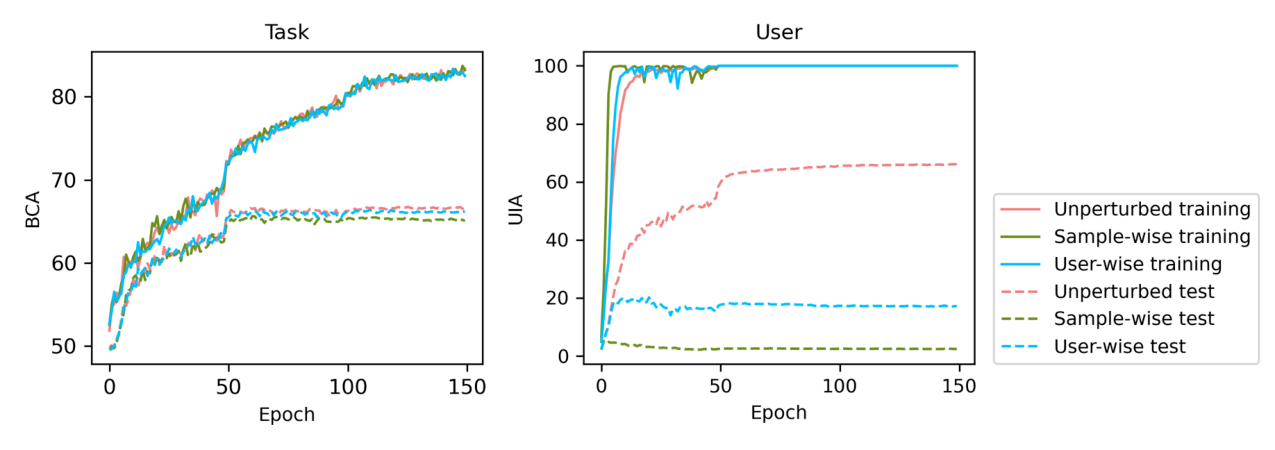}}
\subfigure[]{\label{fig:fig7b}  \includegraphics[width=0.8\linewidth,clip]{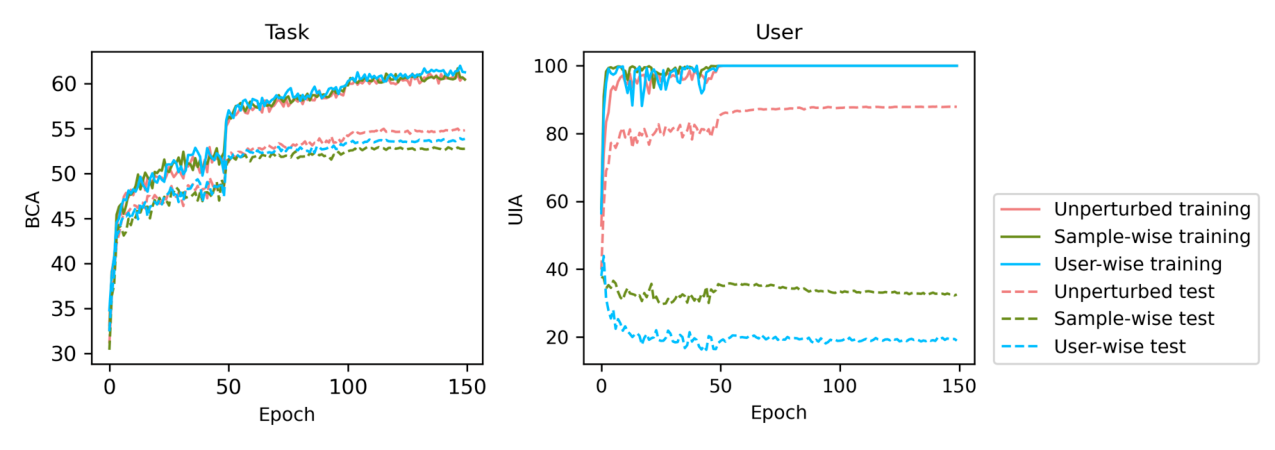}}
\caption{The training and test curves of Task-Classifier and User-Classifier in the training process with the original unperturbed EEG data and the identity-unlearnable EEG data. (a) MI1; and, (b) MI2.} \label{fig:fig7}
\end{figure*}

Figs.~\ref{fig:fig7a} and~\ref{fig:fig7b} show the training and test curves of Task-Classifier and User-Classifier in the training process with the unperturbed data and their perturbed counterparts on the MI1 and MI2 datasets, respectively. The curves on the other five datasets are shown in Supplementary Figs. 17-21. We can observe that:
\begin{enumerate}
	\item The training BCA and UIA curves on the unperturbed data and their perturbed counterparts were very similar, suggesting that it is almost impossible to discover that the data have been perturbed from the training process.
	\item Consistent with Table~\ref{tab:main_result}, the test BCA curves of Task-Classifier trained with the unperturbed and perturbed data were similar; however, the test UIA curves of User-Classifier trained with the identity-unlearnable data dramatically decreased, compared with those on their unperturbed counterparts, indicating that the user identity related information is removed.
\end{enumerate}

\subsection{Privacy protection in online scenario}

We further verified the effectiveness of our proposed approaches in an online scenario, where EEG data of different users are sequentially acquired. We simulated this scenario on the MI1 dataset which has many users, assuming that EEG data of 10 users are collected each time. We converted the newly acquired EEG data into identity-unlearnable ones and added them to the dataset.

\begin{figure}[!t]\centering
\includegraphics[width=0.8\linewidth,clip]{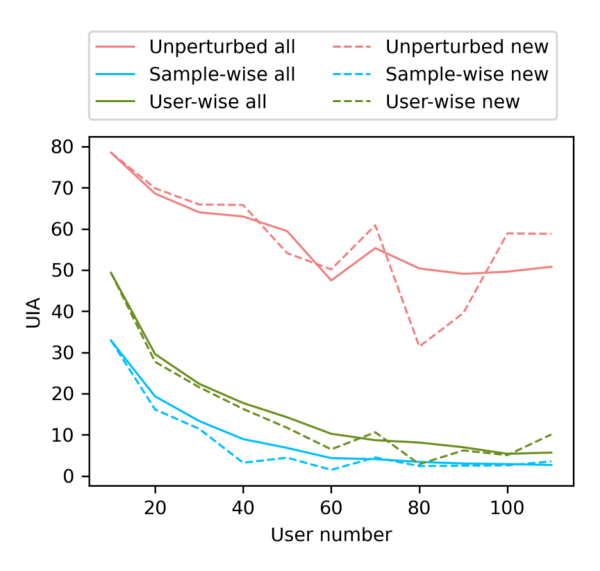}
\caption{UIAs on the EEG data of existing users and new users in a simulated online scenario on MI1.} \label{fig:fig8}
\end{figure}

Fig.~\ref{fig:fig8} shows the UIAs of User-Classifier on existing users and new users. UIAs on the identity-unlearnable data were maintained at a very low level as the number of users increased, indicating the effectiveness of our approaches in online learning, i.e., we can protect the identify privacy of the new users while maintaining the privacy-protection ability for previous users. On the contrary, the UIAs almost always exceeded $50\%$ if the original unperturbed EEG data were used.

\section{Discussions}

This paper has proposed two approaches to perturb the original EEG data into identity-unlearnable EEG data, which protect the user identity privacy while maintaining the primary BCI task classification accuracy. Experiments on seven EEG datasets from five different BCI paradigms demonstrated their effectiveness. Furthermore, we validated the effectiveness of the proposed approaches on the emotion recognition dataset SEED\cite{Zheng2015}. However, considering the relatively low UIA of $34.22\%$ on the unperturbed SEED dataset, we did not include the results on the SEED in the paper. It may be attributed to the limited amount of user data in each session, which leads to overfitting of the models.

In the experiments, we used EEGNet as a feature extractor to generate identity-unlearnable perturbations, and utilized the transferability of the perturbations across models to apply them to DeepCNN and ShallowCNN. However, the transfer effectiveness may depend on the similarity between models, e.g., the greater differences between DeepCNN and EEGNet led to relatively less effective results on DeepCNN.

We further explored the effectiveness of the perturbations generated by EEGNet on Long Short-Term Memory (LSTM). Table~\ref{tab:LSTM} shows the UIAs of the 1D-LSTM model \cite{Sun2019} for user identification on identity-unlearnable EEG data. The average reduction of UIAs is lower than that in Table~\ref{tab:main_result}, indicating again that the transferability of the perturbations is impacted by the similarity between models. Our future research will investigate how to improve the perturbation transferability.

Our proposed approaches remove user identity from the EEG data, but there are also scenarios that user identity could be useful. For example, EEG recognition, or linking the same user's EEGs across different devices. In these scenarios, our approaches can be used for cryptography, i.e., recovering the original EEG data by subtracting the identity-unlearnable perturbations. Additionally, since the identity-unlearnable perturbations are user-specific, the user-wise perturbations can also be used as features for user identification.

\begin{table}[htbp] \centering
\caption{UIAs (\%) of 1D-LSTM on the unperturbed EEG data and those with sample-wise or user-wise perturbations.} \renewcommand\arraystretch{1.1} \setlength{\tabcolsep}{1.8mm}
\begin{tabular}{c|c|c|c|c} \toprule
\multirow{2}[0]{*}{Dataset} & Unperturbed & Sample-wise & Subject-wise & Average \\
& EEG & Perturbation & Perturbation & Reduction \\ \midrule
MI1   & 2.36   & 2.56   & 2.65   & -0.25  \\
MI2   & 51.90  & 47.56  & 37.08  & 9.58  \\
P300  & 73.49  & 56.32  & 14.13  & 38.27  \\
ERN   & 63.84  & 47.68  & 44.58  & 17.71  \\
SSVEP & 8.45   & 9.44   & 9.86   & -1.20  \\
NS    & 95.15  & 93.42  & 9.72   & 43.58  \\
TUSZ  & 91.71  & 80.79  & 60.85  & 20.89  \\ \midrule
Average & 55.27  & 48.25  & 25.55  & 18.37  \\ \bottomrule
\end{tabular}
\label{tab:LSTM}
\end{table}

\section{Conclusions}

EEG signals contain rich private information, e.g., user identity, emotion, and so on, which should be protected in EEG-based BCIs. This paper has exposed a serious privacy problem in EEG-based BCIs, i.e., the user identity in EEG data can be easily learned so that different sessions of EEG data from the same user can be associated together to more reliably mine private information. To address this issue, we proposed two approaches to convert the original EEG data into identity-unlearnable EEG data, i.e., removing the user identity information while maintaining the good performance on the primary BCI task. Experiments on seven EEG datasets from five different BCI paradigms showed that on average the generated identity-unlearnable EEG data can reduce the user identification accuracy by at least 48.65\%, greatly facilitating user privacy protection in EEG-based BCIs.


\end{document}